\documentclass{webofc}
\usepackage[varg]{txfonts}   
\usepackage{amsmath}
\usepackage{bm}

\begin{document}
\title{Diffusion of charm and beauty in the Glasma}
%
%

\author{\firstname{Marco} \lastname{Ruggieri}\inst{1}\fnsep\thanks{\email{ruggieri@lzu.edu.cn}} \and
        \firstname{Santosh Kumar} \lastname{Das}\inst{1,2}}

\institute{School of Nuclear Science and Technology, Lanzhou University, 222 South Tianshui Road,
Lanzhou 730000, China. 
\and
          School of Physical Science, Indian Institute of Technology Goa, Ponda, Goa 403401, India. 
          }

\abstract{%
Relativistic nuclear collisions offer a unique way to study strong interactions at very high energy. The collision process can be described within the gluon saturation framework as the interaction of two colored glasses, and because of this interaction strong longitudinal gluon fields, namely the Glasma, are produced immediately after the collision. Besides, heavy quarks are also produced in the very early stage and because of their large mass and small concentration, their motion does not affect the evolution of the Glasma, thus behaving as ideal probes of the Glasma itself. We study the evolution of the heavy quarks in the Glasma allegedly produced in high energy p-Pb collisions by solving consistently the equations of motion of the quarks in the evolving Glasma fields. We find that this motion can be understood in terms of diffusion in momentum space, similarly to the random motion of a heavy probe in a hot thermalized medium. We 
show how the diffusion of heavy probes affects the nuclear modification factor of D and B mesons in p-Pb collisions. 

}
\maketitle
\section{Introduction}
\label{sec:intro}

The study of the initial condition of the system produced by high energy collisions  
is a difficult but interesting problems related to the physics of
relativistic heavy ion collisions (RHICs), as well as to that of high energy proton-proton (pp) and 
proton-nucleus (pA) collisions. At very high collision energy the two colliding projectiles
can be described within the color-glass-condensate (CGC) effective theory 
\cite{McLerran:1993ni,McLerran:1993ka,McLerran:1994vd}
(see \cite{Gelis:2010nm} for a review and a more complete reference list),
in which fast partons are frozen by time dilation and act like static sources
for low momentum gluons: their large occupation number allows for a classical treatment of these fields.  
The collision of two colored glasses leads to the formation of strong gluon fields in the forward light cone
named the Glasma \cite{Lappi:2006fp};
in weak coupling this consists of longitudinal color-electric and color-magnetic fields
and they are described by the Classical Yang-Mills  (CYM) theory.
Among the high energy collisions mentioned above, pA 
attract a lot of interest because they allow for both a theoretical
and an experimental study of the cold nuclear matter effects (CNME) which are not 
related to the formation of the quark-gluon plasma (QGP): these include shadowing \cite{Eskola:2009uj} and gluon 
saturation \cite{Fujii:2013yja,Ducloue:2015gfa,Rezaeian:2012ye}.

Heavy quarks, namely charm and beauty, are excellent probes of the system created in high energy nuclear collisions, see
\cite{Rapp:2018qla,Aarts:2016hap,Greco:2017rro,Das:2015ana,Das:2013kea,Das:2016cwd,
Das:2017dsh,Das:2015aga,Beraudo:2015wsd,Xu:2015iha,Ozvenchuk:2017ojj,Mrowczynski:2017kso,
Prino:2016cni,Andronic:2015wma} and references therein.
As a matter of fact, their formation time can be roughly estimated as $\tau_\mathrm{form} \approx 1/(2m)$ with
$m$ the quark mass which gives $\tau_\mathrm{form} \leq 0.1$ fm/c for charm and  $\tau_\mathrm{form} \leq 0.04$ fm/c
for beauty. Moreover, their number is very small compared to that of light quarks, 
thus charm and beauty carry a very tiny color current and their effect on the background evolving 
bulk (light quarks and gluons) is negligible. Because of these, charm and beauty can propagate and probe the
evolving Glasma produced in high energy collisions bringing no disturbance to it.

In this proceeding we report on our recent study \cite{Ruggieri:2018rzi} in which we 
have focused on the diffusion of charm in high energy p-Pb collisions,
and we extend the aforementioned study by publishing for the first time a result obtained
for the diffusion of beauty.
The main purpose of our study is to
compute consistenly the propagation of the heavy quarks in the initial gluon fields,
then computing the nuclear modification factor, $R_{\mathrm{pPb}}$,
for $D-$mesons and $B-$mesons that has
been measured recently~\cite{Abelev:2014hha, Aaij:2017gcy}. In fact, in \cite{Ruggieri:2018rzi} 
we have found that the propagation charm in the  classical gluon field leads to $R_{\mathrm{pPb}}$
that reminds qualitatively the one measured for $D-$mesons in p-Pb collisions.

Because in our study we do not have a longitudinal expansion we do not attempt to a direct 
quantitative comparison with existing experimental data: while this will be the subject of forthcoming
publications, we feel it is very important to understand the theoretical problem of the diffusion
of heavy colored probes in the Glasma which has been overlooked by the community,
and having this purpose in mind a first necessary step is to understand quantitatively
the diffusion of these probes in a static gluon medium. 
It is worth mentioning that propagation of heavy quarks in the Glasma has been studied 
for the first time in \cite{Mrowczynski:2017kso}
within a Fokker-Planck equation.
The main differences with respect to \cite{Mrowczynski:2017kso} are that we do not rely on the small
transferred momentum expansion, and we include the dynamical
evolution of the gluon medium, eventually
offering also a link to observables.
Within our framework we solve consistently the CYM equations and the Wong equations; 
this approach is equivalent to solve the Boltzmann-Vlasov equations
for the heavy quarks in a collisionless plasma.
Nevertheless, our main results do not differ drastically by those of  \cite{Mrowczynski:2017kso}
and we agree with that reference by describing the evolution of the heavy probes in terms of 
diffusion in momentum space, similarly to what happens in a thermal medium.

\section{The model}
\label{sec:model}
 
In this section we briefly describe the model we implement in our calculations;
more details can be found in \cite{Ruggieri:2018rzi}.
For the distribution of the color charges in the two colliding objects we adopt the standard MV model, 
in which the static color charge densities $\rho_a$   
on the nucleus $A$ are assumed to be random variables that are 
normally distributed with zero mean 
and variance specified by the equation
\begin{equation}
\langle \rho^a_A(\bm x_T)\rho^b_A(\bm y_T)\rangle = 
(g^2\mu_A)^2 \varphi_A
\left(
\bm X_T
\right)\delta^{ab}\delta^{(2)}(\bm x_T-\bm y_T);
\label{eq:dfg}
\end{equation}
here $A$ corresponds to either the proton or the Pb nucleus, $a$ and $b$ denote the adjoint color index;
in this work we limit ourselves for simplicity to the case of the $SU(2)$ color group therefore
$a,b=1,2,3$.
In Eq.~(\ref{eq:dfg}) $g^2\mu_A$ denotes the color charge density and it is of the order of the saturation 
momentum $Q_s$ \cite{Lappi:2007ku}.
 
The function $\varphi_A(\bm X_T)$  in Eq.~\eqref{eq:dfg}
with $\bm X_T = (\bm x_T + \bm y_T)/2$ allows for a nonuniform probability distribution
of the color charge in the transverse plane.
For the case of the Pb nucleus in a p-Pb collision we assume a uniform 
probability and take $\varphi(\bm x_T)=1$.  
On the other hand, for the proton we distribute the color charges by virtue of a standard gaussian
distribution \cite{Schenke:2014zha,Schenke:2015aqa,Mantysaari:2017cni,Mantysaari:2016jaz}: 
\begin{equation}
\varphi_p(\bm X_T)=e^{-(\bm X_T^2)/(2 B_{cq})},\label{eq:xi_proton}
\end{equation} 
with 
$B_{cq} = 3$ GeV.
For the Pb nucleus we limit ourselves to show results obtained for $g^2\mu_\mathrm{Pb} = 3.4$ GeV while for the proton
we use $g^2\mu_\mathrm{p} =1.41$ GeV: a detailed discussion on how these parameters are fixed can be found in  \cite{Ruggieri:2018rzi}.

In order to determine the Glasma fields at $t=0^+$
we  firstly solve the Poisson equations for the gauge potentials, namely
\begin{equation}
-\partial_\perp^2 \Lambda^{(A)}(\bm x_T) = \rho^{(A)}(\bm x_T).
\end{equation}
Wilson lines are computed as
$
V^\dagger(\bm x_T) = e^{i \Lambda^{(A)}(\bm x_T)}$, 
$W^\dagger(\bm x_T) = e^{i \Lambda^{(B)}(\bm x_T)}$,
and the pure gauge fields of the two colliding nuclei are given by
$
\alpha_i^{(A)} = i V \partial_i V^\dagger$,
$\alpha_i^{(B)} = i W \partial_i W^\dagger$.
In terms of these fields the solution of the CYM in the forward light cone
at initial time, namely the Glasma potentials, 
are given by
$A_i = \alpha_i^{(A)} + \alpha_i^{(B)}$~ for $i=x,y$ and $A_z = 0$,
and the initial Glasma color-electric and color-magnetic fields are 
\begin{eqnarray}
&& E^z = i\sum_{i=x,y}\left[\alpha_i^{(B)},\alpha_i^{(A)}\right], \label{eq:f1}\\
&& B^z = i\left(
\left[\alpha_x^{(B)},\alpha_y^{(A)}\right]  + \left[\alpha_x^{(A)},\alpha_y^{(B)}\right]  
\right),\label{eq:f2}
\end{eqnarray}
while the transverse fields are vanishing. 
 
The dynamical evolution of Glasma is given by the 
CYM equations, namely
\begin{eqnarray}
\frac{dA_i^a(x)}{dt} &=& E_i^a(x),\\
\frac{dE_i^a(x)}{dt} &=& \sum_j \partial_j F_{ji}^a(x) + 
\sum_{b,c,j} f^{abc} A_j^b(x)  F_{ji}^c(x).\label{eq:CYM_el}
\end{eqnarray}
where the magnetic part of the field strength tensor is  
\begin{equation}
F_{ij}^a(x) = \partial_i A_j^a(x) - \partial_j A_i^a(x)  + \sum_{b,c}f^{abc} A_i^b(x) A_j^c(x);
\label{eq:Fij}
\end{equation}
here $f^{abc} = \varepsilon^{abc}$ with $\varepsilon^{123} = +1$.

We initialize the heavy probes in the transverse momentum plane by means of the standard
Fixed Order + Next-to-Leading Log (FONLL) QCD calculation which 
describes the D-mesons spectra in $pp$ collisions after fragmentation~\cite{FONLL, Cacciari:2012ny,Cacciari:2015fta} 
\begin{equation}
\left.\frac{dN}{d^2 p_T}\right|_\mathrm{prompt} = \frac{x_0}{(x_1 + p_T)^{x_2}};\label{eq:HQ_1}
\end{equation}
a similar equation holds for beauty. 
We also assume that the initial longitudinal momentum vanishes.
Initialization in coordinate space is done as follows:
the tranverse coordinates distribution is built up by means of the
function $\varphi(\bm x_T)$ in Eq.~\eqref{eq:xi_proton}, because we expect the
heavy quarks to be produced in the overlap region of proton and Pb nucleus that
coincides with the transverse area of the proton;
on the other hand, we use a uniform distribution for the longitudinal coordinate.

The dynamics of heavy quarks in the Glasma is studied by the Wong equations  \cite{Wong:1970fu} that
for a single quark can be written as
\begin{eqnarray}
\frac{d x_i}{dt}& =& \frac{p_i}{E},\\
E\frac{d p_i}{dt} &=& Q_a F^a_{i\nu}p^\nu,\\
E\frac{d Q_a}{dt} &=& - Q_c\varepsilon^{cba} \bm A_b\cdot\bm p,
\end{eqnarray}
where $i=x,y,z$; here, the first two equations are the familiar Hamilton equations of motion for the coordinate and its conjugate
momentum, while  the third equation corresponds to the gauge invariant color current conservation.
Here $E = \sqrt{\bm p^2 + m^2}$ with $m=1.5$ GeV for charm and $m=4.5$ GeV for beauty. 
 More details can be found in  \cite{Ruggieri:2018rzi}.
 
\section{Results}
\label{sec:results}

\begin{figure*}[t!]
\begin{center}
\includegraphics[width=6cm]{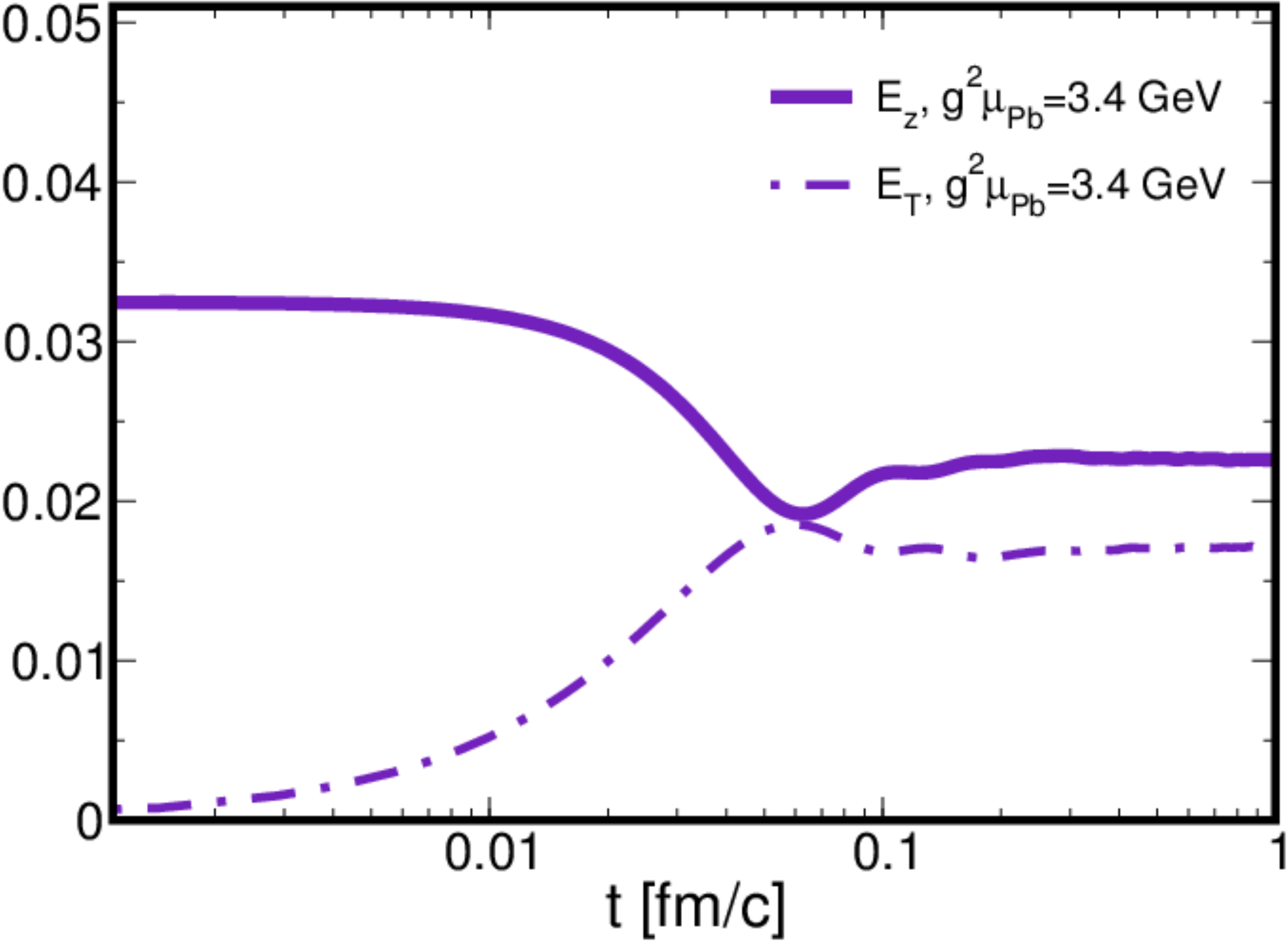}
\caption{\label{fig-0}    Averaged color-electric fields for p-Pb collision, measured in lattice units.
Solid line corresponds to the longitudinal field while dashed line denote the transverse fields;
we have put  $g^2\mu_\mathrm{Pb}=3.4$ GeV. Lattice spacing is $\delta x=0.04$ fm.
Adapted from \cite{Ruggieri:2018rzi}.}   
\end{center}
\end{figure*}

In Fig.~\ref{fig-0} we plot the averaged color-electric fields as a function of time.
Solid line corresponds to the longitudinal field while dashed line denote the transverse fields;
we have put  $g^2\mu_\mathrm{Pb}=3.4$ GeV and lattice spacing is $\delta x=0.04$ fm \cite{Ruggieri:2018rzi}.
At initial time the system consists of purely longitudinal fields but the gluon dynamics leads to a quick production
of transverse fields: in fact, within $t\approx 0.1$ fm/c a bulk made of both longitudinal and transverse fields is formed.

\begin{figure*}[t!]
\begin{center}
\includegraphics[width=6cm]{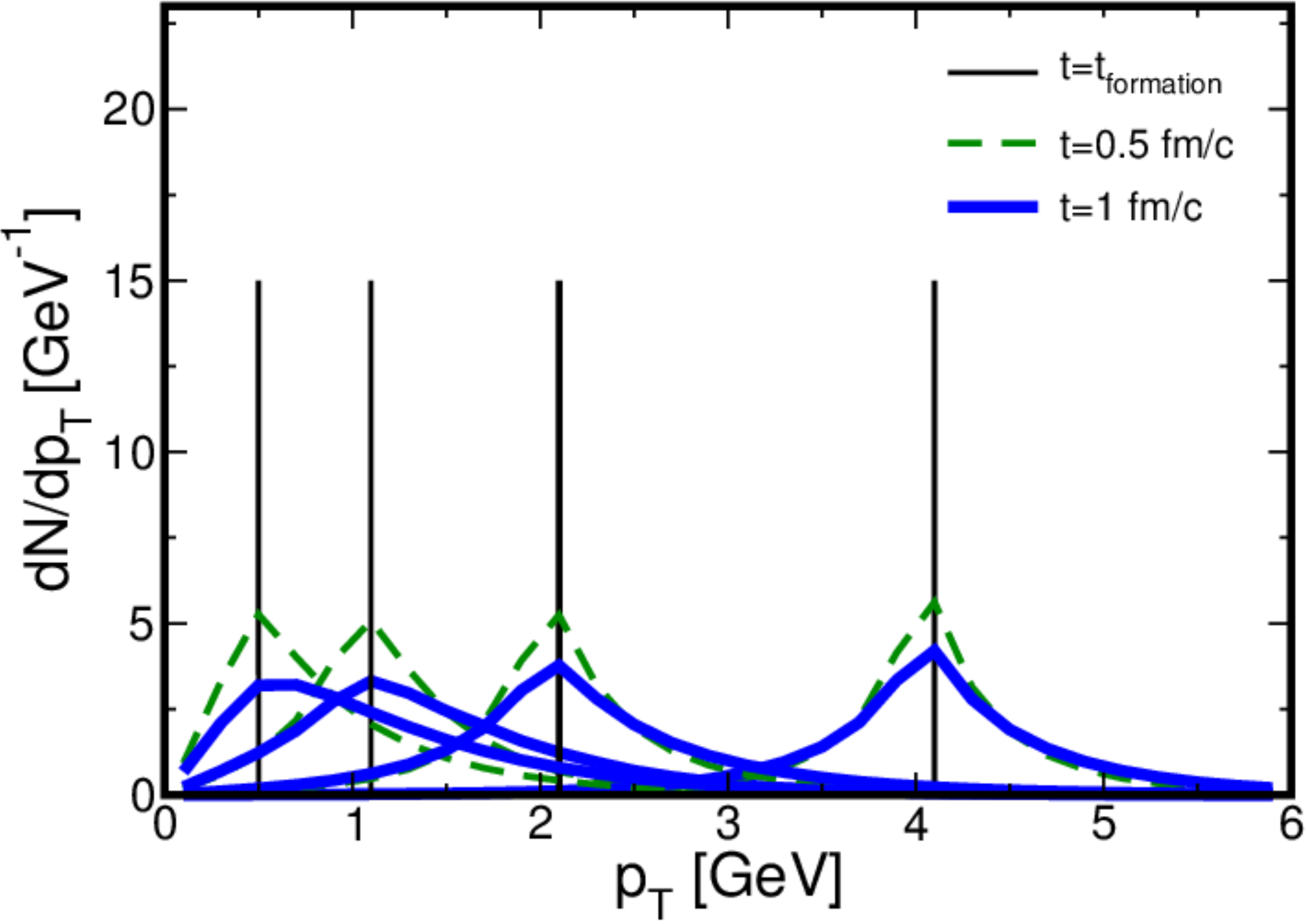}
\caption{\label{fig-1}    Evolution of $\delta-$distribution functions
of $c-$quarks in the Glasma fields. Black solid lines correspond to the initializations,
green dashed lines to $t=0.5$ fm/c and
blue solid lines to $t=1$ fm/c. We take $g^2\mu_\mathrm{Pb} = 3.4$ GeV.
Adapted from \cite{Ruggieri:2018rzi}.}   
\end{center}
\end{figure*}

To elucidate the interaction of the $c-$quarks with the evolving Glasma 
we prepare initializations in which we put all the $c-$quarks in a thin $p_T$ bin to obtain a $\delta-$like distribution;
the evolution of this distribution is studied by means of the Wong equations.
This study is useful to better understand the interaction of the Glasma with different $p_T$ modes.
The results of this are shown in Fig.~\ref{fig-1}  in which we plot the spectrum of charms
at initial time (solid black lines), at $t=0.5$ fm/c (green dashed lines)
and at $t=1$ fm/c (solid blue lines) for several values of the initial $p_T$.
We notice that in all the cases examined here the interaction with the Glasma
leads to the spreading of the spectrum, analogously to the standard diffusion in momentum space
encountered in a Brownian motion.
In addition to this, for low $p_T$ we find a drift towards higher momenta
that can be described as an average acceleration of the $c-$quarks.
We notice that a drift is also present for higher values of $p_T$ and tends to lower
the $\langle p_T\rangle$ despite this effect is very small and for large $p_T$ the diffusion dominates
over the drift.

\begin{figure*}[t!]
\begin{center}
\includegraphics[width=6cm]{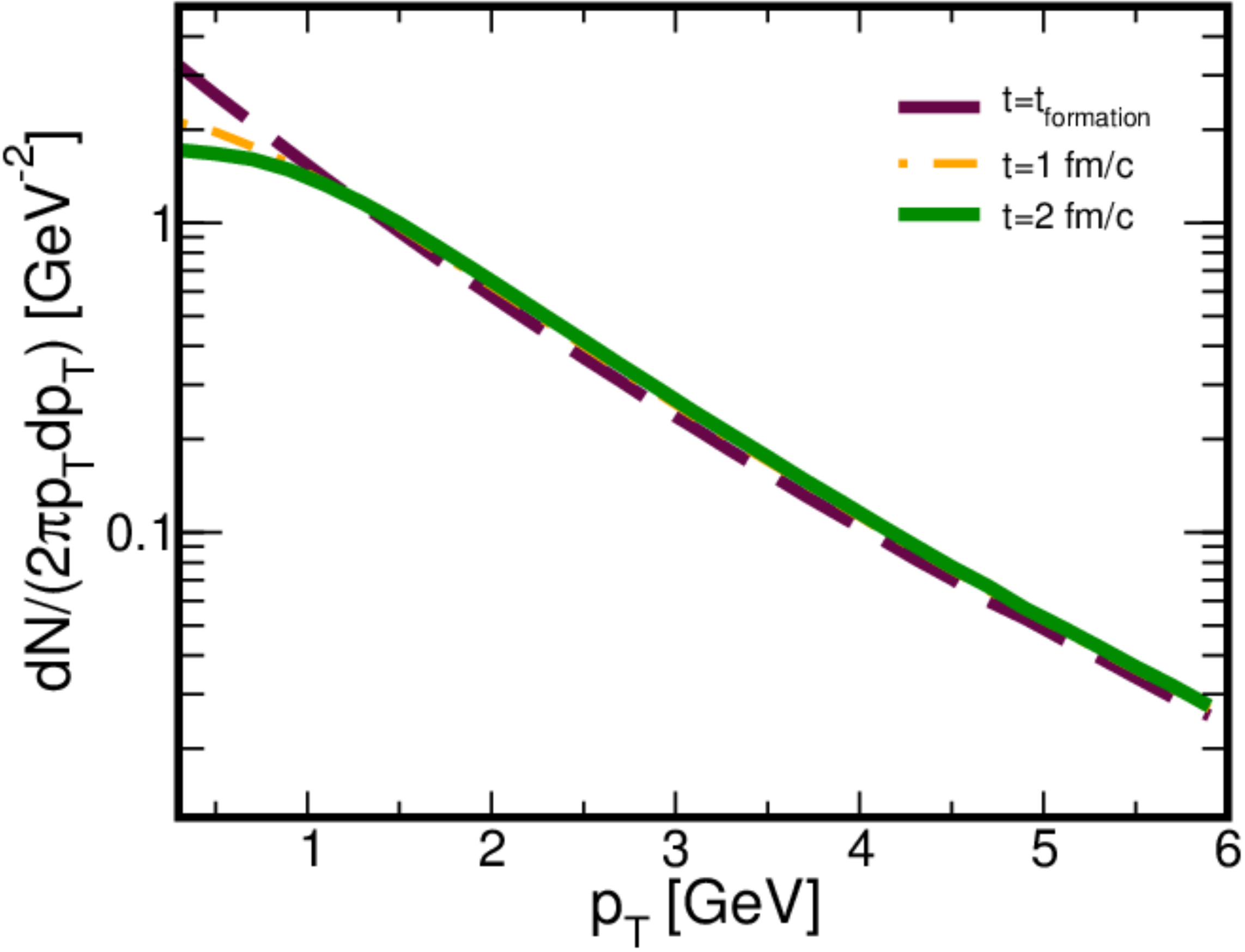}
\caption{\label{fig-2}  Spectrum of $c-$quarks
at the initial time (dashed maroon line), at $t=1$ fm/c (orange dot-dashed line) and
at $t=2$ fm/c (green solid line). We have put $g^2\mu_\mathrm{Pb} = 3.4$ GeV.}   
\end{center}
\end{figure*}

In Fig.~\ref{fig-2} we plot the evolution of the charm spectrum with time. At the initial time the spectrum is given by the
standard FONLL form; the interaction with the evolving Glasma bends downwards the spectrum at low $p_T$ as a result of the diffusion
discussed above: particles migrate from low to high $p_T$ similarly to the acceleration that charged particles would experience 
in a strong electric field.

\begin{figure*}[t!]
\begin{center}
\includegraphics[width=8cm]{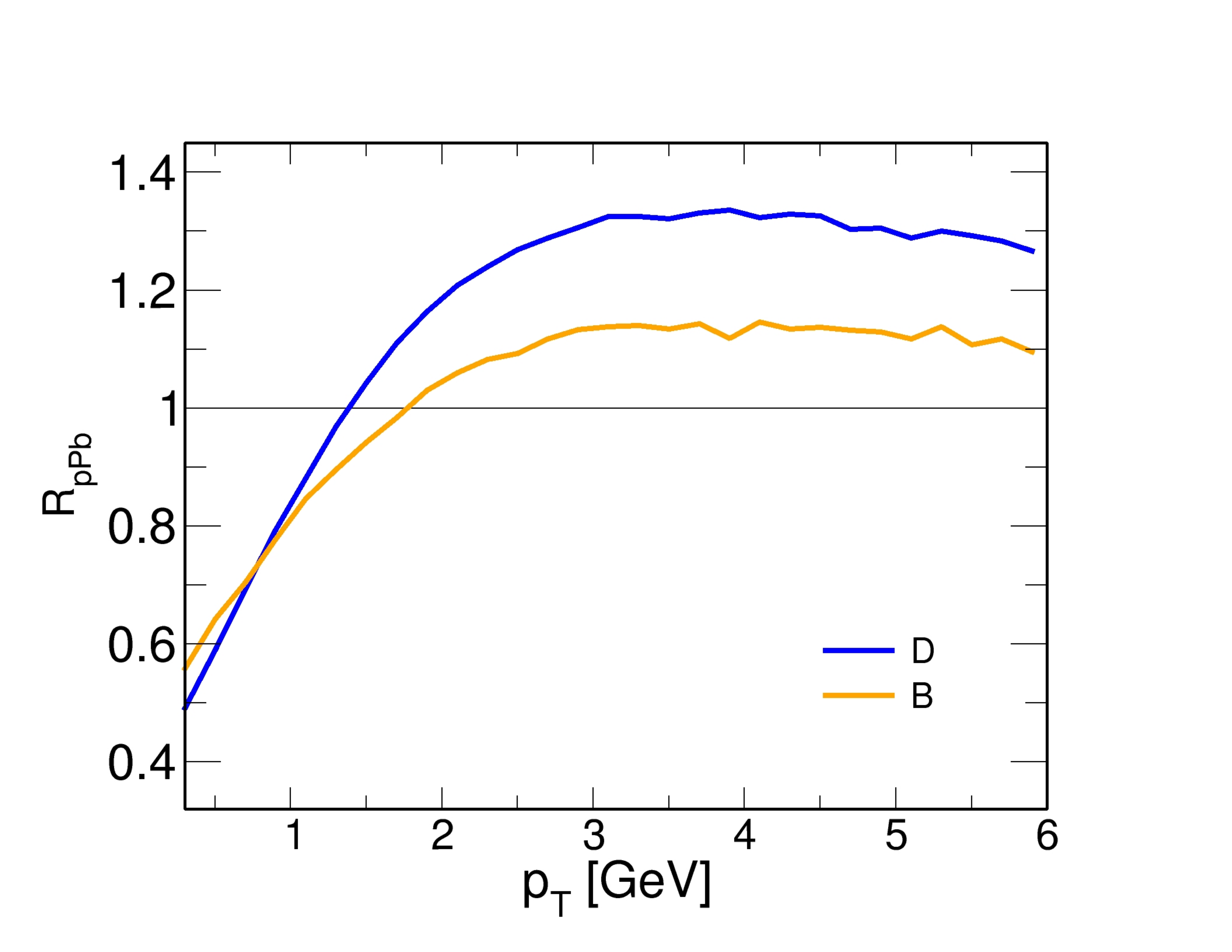}
\caption{\label{fig-3}    Nuclear modification factor for $D-$mesons (blue line) and $B-$mesons (orange line).}   
\end{center}
\end{figure*}

In Fig.~\ref{fig-3} we plot the nuclear modification factor $R_\mathrm{pPb}$ defined as
\begin{equation}
R_\mathrm{pPb} = \frac{\left(dN/d^2 p_T\right)_\mathrm{Glasma}}{\left(dN/d^2 p_T\right)_\mathrm{FONLL}},
\label{eq:llp}
\end{equation}
where $\left(dN/d^2 p_T\right)_\mathrm{Glasma}$ corresponds to the spectrum computed after the evolution of the
heavy quarks with the Glasma: blue line denotes  $R_\mathrm{pPb}$ for $D-$mesons and
orange line corresponds to the same quantity for $B-$mesons. For both the lines we have used the standard
fragmentation framework to extract the momentum distribution of mesons from the underlying distributions 
of quarks \cite{Pet}. For both charm and beauty we have stopped the calculation
at $t=1$ fm/c. We notice that because of the diffusion in momentum space described above,
the  $R_\mathrm{pPb}$ is smaller than one up to $p_T\approx 1$ GeV; on the other hand, for large values of $p_T$
it remains consistend with the perturbative FONLL within a $20\%$.
The deviation of $R_\mathrm{pPb}$ from one is a direct consequence of the interaction of the heavy probes
with the Glasma; this effect is more pronounced for the $c-$quarks while for beauty the overall effect is much milder
because of its larger mass.
Since in our calculation the longitudinal expansion is missing, we do not compare our results
with experimental data \cite{Abelev:2014hha, Aaij:2017gcy}, despite we notice that the qualitative shape of these
agrees with our theoretical calculations: in the future we will  upgrade our calculations in order to take the
longitudinal expansion into account in order to produce results that can be directly compared to experimental data.

\section{Conclusions}
\label{sec:conclusion}

In this talk we have reported on our study about the diffusion of heavy probes, namely charm and beauty,
in the early stage of high energy nuclear collisions, focusing on the case of pA collisions.
The dynamics of the gluon fields has been studied by means of the classical Yang-Mills equations; that of the
heavy probes has been studied by the Wong equations solved on the top of the evolving background gluon field. 

We have found that the evolution of the heavy quarks in the Glasma can be understood
in terms of diffusion in momentum space \cite{Ruggieri:2018rzi} in agreement with \cite{Mrowczynski:2017kso}.
This diffusion leads to a nuclear modification factor $R_\mathrm{pPb}$ different from one, signaling
the effect of the interactions with the gluon fields during the evolution. In particular, for low $p_T$
we have found $R_\mathrm{pPb}\leq 1$ as a result of the diffusion and drift of small energy quarks towards
higher momenta, similarly to what would happen to probe electric charges in a strong electric field.
We have computed the nuclear modification factor for both $D-$mesons and $B-$mesons,
studying the diffusion of both charm and beauty with the background gluon field.
The diffusion is qualitatively the same for both charm and beauty, although quantitatively
beauty experiences a smaller effect than charm because of its larger mass.

We have not attempted a direct comparison with experimental data because our calculations miss
the longitudinal expansion; nevertheless, we feel that our results are an important step towards a more complete
understanding of the dynamics of heavy colored probes in the strong gluon fields produced in the early stages
of high energy nuclear collisions, and suggest that this dynamics should be further studied in order to 
have a quantitative estimate of the role of the pre-thermalization epoch on heavy quark observables
in pp, pA and AA collisions.

{\it Acknowledgements.} M. R. acknowledges the committee of QCD@Work2018 held in Matera (Italy)
where this talk has been presented, for their kind ospitality and the beautiful and inspiring environment that they have built up
during the workshop, assisted by the marvelous architecture and breathtaking landscapes of the town where the workshop has
taken place.
The work of the authors is supported by the National Science Foundation of China (grants number 11875153
and 11805087)
and by the Fundamental Research Funds for the Central Universities (grant number 862946).

\end{document}